\documentclass[twocolumn,showpacs,prd,floatfix,axodraw]{revtex4}
\usepackage{mathrsfs}
\usepackage{graphicx,,booktabs,bm}
\usepackage{overpic}
\usepackage{color}
\usepackage{amssymb}

\usepackage{mathrsfs,bm,amsmath,amssymb}
\usepackage{longtable,lscape}
\usepackage{txfonts}
\usepackage{amssymb}
\usepackage{indentfirst}
\usepackage{graphicx,,booktabs}
\usepackage{multirow,ulem}

\usepackage{color}
\usepackage{amssymb}

\definecolor{cover}{rgb}{0.77,0.87,0.88}
\definecolor{blueone}{rgb}{0.1,0.1,.7}
\definecolor{citec}{rgb}{0.14,0.47,0.09}
\definecolor{two}{rgb}{0.0,0.5,0.}
\definecolor{three}{rgb}{.5,.1,0.15}
\usepackage[bookmarks=true,bookmarksopen=false,plainpages=false,breaklinks=true,
   bookmarksnumbered=true,hypertexnames=false,
   filecolor=blue,urlcolor=three,menucolor=three,
   linkcolor=three,citecolor=blueone, colorlinks,
   anchorcolor=blue,runcolor=pink,frenchlinks=red
   pdfstartview=FitH,pdftitle=title,%
   pdfauthor=author]{hyperref}

\begin{document}
\title{Possible $P$- wave $D_s\bar{D}_{s0}(2317)$ molecular state $Y^{'}(4274)$}

\author{Hong Qiang Zhu}
\affiliation{College of Physics and Electronic Engineering, Chongqing Normal University, Chongqing 401331,China}

\author{Yin Huang\footnote{corresponding author}} \email{yin.huang@apctp.org}
\affiliation{Asia Pacific Center for Theoretical Physics,
Pohang University of Science and Technology, Pohang 37673, Gyeongsangbuk-do,
South Korea}
\affiliation{School of Physical Science and Technology, Southwest Jiaotong University, Chengdu 610031,China}

\begin{abstract}
Stimulated by the measurement of the $J\psi\phi$ decay model of $Y(4274)$ by the LHCb Collaboration, we consider a possible interpretation
of this state as a hadron molecular-a bound state of $D_s$ and $\bar{D}_{s0}(2317)$ mesons.  Using effective Lagrangian approach, we calculate
the two-body strong decay channels $Y(4274)\to{}J/\psi\phi,\chi_{c0}\eta,\chi_{c0}\eta,D^{*}_s\bar{D}_s$, $D\bar{D}^{*}$, $K\bar{K}^{*}$, and $\phi\phi$
through hadronic loops and three-body decays into $\pi^0{}D_s\bar{D}_s$.  In comparison with the LHCb data, our results show that $Y(4274)$
cannot be assigned to be a $D_s\bar{D}_{s0}(2317)$ molecular state.  The calculated partial decay widths with $J^P=1^{+}$ $D_s\bar{D}_{s0}$
molecular state picture indicate that allowed decay modes, $\chi_{c0}\eta$ and $\chi_{c1}\eta$, may have the smallest branching ratio and are
of the order of 0.0 MeV.  Future experimental measurements of such two processes can be quite useful to test the different interpretations of
the $Y(4274)$. If $P-$wave $D_s\bar{D}_{s0}$ molecular exist [we marked as $Y^{'}(4274)$],  the total decay is at the order of 1.06-1.84 MeV,
which seems to be within the reach of the current experiments such as Belle II.  In addition, the calculated partial decay widths indicate that
allowed decay mode, $D\bar{D}^{*}$, may have the biggest branching ratio.  The experimental measurements for this strong decay process could be
a crucial to observe such a new state $Y^{'}(4274)$.
\end{abstract}

\date{\today}


\maketitle
\section{Introduction}
Thanks to the great progress of the experiment in the past several decades,  many hadrons that cannot be ascribed into
the simple $\bar{q}q$ configuration for mesons or $qqq$ configuration for baryons have been reported~\cite{Zyla:2020zbs}.
For example, various hidden-charm pentaquark were observed in the $J/\psi{}p(\Lambda)$ invariant mass from the heavy
baryon decay $\Lambda_b^{+}\to{}K^{-}J/\psi{}p$ ($\Xi_b^{-}\to{}K^{-}J/\psi{}\Lambda$) at LHCb Collaboration, $P_c(4312,4440,4450)$
and $P_{cs}(4459)$~\cite{Aaij:2019vzc,LHCb:2020jpq}.
Their confirmation, and determination of their quantum numbers, would allow new insights into the binding mechanisms
present in multiquark systems, and help improve understanding of QCD in the nonperturbative regime.

In 2017, a charmonium-like meson named $Y(4274)$ was observed again by the LHCb collaboration in the analysis of the
$B^{+}\to{}J/\psi\phi{}K^{+}$ reaction~\cite{LHCb:2016axx,LHCb:2016nsl}.   The observed resonance masses, widths,
and favorable quantum numbers are
\begin{align}
M&=4273.3\pm{}8.3^{+17.2}_{-3.6}~~~{\rm MeV},\nonumber\\
\Gamma&=56\pm11^{+8.0}_{-11}~~~~~ {\rm MeV},~~J^{PC}=1^{++},
\end{align}
respectively,  which are consistent with the early CDF Collaboration~\cite{Yi:2010aa} report
\begin{align}
M&=4274.4^{+8.4}_{-6.7} (stat)~~~{\rm MeV},\nonumber\\
\Gamma&=32.3^{+21.9}_{-15.3} (stat)~~~~~ {\rm MeV}.
\end{align}
Since the statistic is not enough, its spin-parity quantum number was not confirmed by the CDF Collaboration.
However, the isospin of this state is zero and it contains at least four valence quarks from the observed
$J/\psi{}\phi$ decay mode.

Following the discovery of the $Y(4274)$, several theoretical studies have been performed.  In the QCD sum rules approach,
based on the analysis of the mass spectrum, the $Y(4274)$ can be interpreted as the $S$-wave $cs\bar{c}\bar{s}$ state with
spin-parity $J^P = 1^+$ ~\cite{Chen:2016oma,Wang:2016dcb}.  The compact tetraquark model, implemented by Stancu, can also
describe the $Y(4274)$~\cite{Stancu:2009ka}, while only one $J^P = 1^+$ state exists.  While in Ref.~\cite{Lu:2016cwr} the
mass of the $Y(4274)$ was studied in the relativized quark model and it was shown that the $Y(4274)$ cannot be explained as
a tetraquark state, however, it can be a good candidate of the conventional $\chi_{c1}(3^{3}P_1)$ state.  And the $\chi_{c1}(3^{3}P_1)$
explain for $Y(4274)$ was also proposed in Ref.~\cite{Gui:2018rvv}. In the context of the QCD two-point sum rule method by
taking into account the quark, gluon, and mixed vacuum condensates, Ref.~\cite{Agaev:2017foq} assigned the $Y(4274)$ as a sextet
[$cs$][$\bar{c}\bar{s}$] diquark-antidiquark state with spin-parity $J^P = 1^+$.   Based on the spin-spin interaction,
Maiani et al. suggest that the $Y(4274)$ may have quantum number $0^{++}$ or $2^{++}$~\cite{Maiani:2016wlq},  which contradicts
the experimental observation~\cite{LHCb:2016axx,LHCb:2016nsl}.   Moreover, a detailed calculation is performed by Zhu~\cite{Zhu:2016arf}
where the $Y(4274)$ may be described simultaneously by adding the up and down quark components.

Although the studies of Refs.~\cite{Chen:2016oma,Wang:2016dcb,Stancu:2009ka,Agaev:2017foq,Lu:2016cwr,Gui:2018rvv,Maiani:2016wlq,Zhu:2016arf}
seem to indicate that this state is a compact tetraquark state or diquark-antidiquark state, $Y(4274)$ might still be a $D_s\bar{D}_{s0}(2317)$
hadronic molecule state.  Since the mass of  $Y(4274)$ is about 12 MeV below the threshold of $D_s\bar{D}_{s0}$ ($m_{D_{s0}}=2317.8\pm{}0.6$
MeV and $m_{D_s}=1968.34\pm{}0.07$ MeV~\cite{Zyla:2020zbs}), it is reasonable to regard it is a bound state of $D_s\bar{D}_{s0}$.
The idea comes from molecular state interpretation of deuteron due to deuteron mass is a little below the corresponding threshold and
exhibit sizable spatial extension.  And because the quantum numbers of $\bar{D}_{s0}$ and $D_s$ are $J^P=0^{+}$ and $J^P=0^{-}$, respectively,
to form a bound state with quantum number $J^P=1^{+}$, the coupling between $Y(4274)$ and its constituents should be a $P$ wave.  Indeed,
it is shown in Ref.~\cite{He:2016pfa} that the interaction between a $D_s$ meson and a $\bar{D}_{s0}$ meson is strong enough to form a
bound state with a mass about 4274 MeV.

From Ref~\cite{He:2016pfa}, $Y(4274)$ may be a molecular state.  However, currently, we cannot fully exclude other possible explanations
such as a compact pentaquark state~\cite{Chen:2016oma,Wang:2016dcb,Stancu:2009ka,Agaev:2017foq,Lu:2016cwr,Gui:2018rvv,Maiani:2016wlq,Zhu:2016arf}.
Further research is required to distinguish whether it is a molecular or compact multi-quark state.   One way to distinguish the two scenarios is
to study the allowing strong decay widths of the $Y(4274)$ baryon due to the strong decay almost saturates the total strong decay width.
In the present paper we consider possible strong decay modes using an effective Lagrangian approach by
assuming that $Y(4274)$ is a hadronic molecule state of $D_s$ and $\bar{D}_{s0}$.

This work is organized as follows. The theoretical
formalism is explained in Sec. II. The predicted partial
decay widths are presented in Sec. III, followed by a short summary in the last section.

\section{FORMALISM AND INGREDIENTS}
Besides the $J/\psi{}\phi$ decay model, which other decay is allowed.  We first find that the transition from $Y(4274)$ to final states
composed of purely neutral state $A\bar{A}$ are strictly forbidden by the conservation of the $c$ parity.  Thus, the decay of $Y(4274)$
into $\eta_c\eta$, $\chi_{c0}\eta$, $\chi_{c1}\eta$, $\phi\phi$, $\eta\eta$, $D^{*}_s\bar{D}_s$, $D\bar{D}^{*}$, $K\bar{K}^{*}$ and $\pi^0\bar{D}_sD_s$
are allowed by considering appropriate phase space~\cite{Zyla:2020zbs}.  However, the transitions $Y(4274)\to{}\eta_c\eta$ and $\eta\eta$ are
also strictly forbidden by the conservation of angular momentum.  In this work, we will calculate $J/\psi{}\phi$, $\chi_{c0}\eta$, $\chi_{c1}\eta$,
$D^{*}_s\bar{D}_s$, $D\bar{D}^{*}$, $K\bar{K}^{*}$ and $\phi\phi$ strong decay patterns of $p$-wave $D_s\bar{D}_{s0}$ molecular state within
the effective Lagrangians approach, and find the relation between $D_s\bar{D}_{s0}$ molecular state and $Y(4274)$ by comparing with the experiment
observation.

Before introducing the theoretical framework, we need construct the flavor functions for $D_s\bar{D}_{s0}$ system with definite $I(J^{P{\cal{C}}})$.
Since the $Y(4274)$ carry quantum numbers $I(J^{P{\cal{C}}})=0(1^{++})$, the flavor function for a definite charge parity ${\cal{C}}=1$ can be easy
obtained~\cite{Liu:2013rxa}
\begin{align}
|D_s\bar{D}_{s0}\rangle=\frac{1}{\sqrt{2}}[D_s^{+}D_{s0}^{-}-D_s^{-}D_{s0}^{+}]\label{eqr3}.
\end{align}

Considering the quantum number $J^P=1^{+}$ and the flavor function, $Y(4274)$ should couple to its components dominantly via
$P-$ wave, and the corresponding effective Lagrangian is in the form~\cite{Ma:2010xx}
\begin{align}
{\cal{L}}_{Y(4274)}(x)&=g_{YD_s\bar{D}_{s0}}Y^{\mu}(x)\int{}d^4y\Phi(y^2)\nonumber\\
                      &\times{}\frac{1}{\sqrt{2}}[D^{+}_s(x+\omega_{D^{-}_{s0}}y)\overleftrightarrow{\partial_{\mu}}D^{-}_{s0}(x-\omega_{D^{+}_{s}}y)\nonumber\\
                      &-D^{-}_s(x+\omega_{D^{+}_{s0}}y)\overleftrightarrow{\partial_{\mu}}D^{+}_{s0}(x-\omega_{D^{-}_{s}}y)]\label{eq1},
\end{align}
where $\omega_{D_s}=m_{D_s}/(m_{D_s}+m_{D_{s0}})$ and $\omega_{D_{s0}}=m_{D_{s0}}/(m_{D_s}+m_{D_{s0}})$.
In the Lagrangian, an effective correlation function $\Phi(y^2)$ is introduced to describe the distribution
of the two constituents, the $D_s$ and the $\bar{D}_{s0}$, in the hadronic molecular $Y(4274)$ state.
The introduced correlation function also makes the Feynman diagrams ultraviolet finite. Here we choose the
Fourier transformation of the correlation to be a Gaussian form in the Euclidean space~\cite{Faessler:2007gv,Faessler:2007us,Dong:2008gb,Dong:2009uf,Dong:2009yp,Dong:2017rmg,Dong:2014ksa,Dong:2014zka,Dong:2013kta,Dong:2013iqa,Dong:2013rsa,Dong:2012hc,Dong:2011ys,Dong:2010xv,Dong:2010gu,Dong:2009tg,Dong:2017gaw,Yang:2021pio,Zhu:2020jke} ,
\begin{align}
\Phi(p^2)\doteq\exp(-p_E^2/\Lambda^2)
\end{align}
with $\Lambda$ being the size parameter which characterizes the distribution of the components inside the molecule.
The value of $\Lambda$ could not be determined from first principles, therefore it should better be determined by
experimental data.  It is usually chosen to be about $1$ GeV, which depends on experimental total widths~\cite{Faessler:2007gv,Faessler:2007us,Dong:2008gb,Dong:2009uf,Dong:2009yp,Dong:2017rmg,Dong:2014ksa,Dong:2014zka,Dong:2013kta,Dong:2013iqa,Dong:2013rsa,Dong:2012hc,Dong:2011ys,Dong:2010xv,Dong:2010gu,Dong:2009tg,Dong:2017gaw,Yang:2021pio,Zhu:2020jke}. In this work, we vary  $\Lambda$ in a range of 0.9 GeV $\leq\Lambda\leq{}1.10$ GeV to study whether the $Y(4274)$
can be interpreted as $P-$ wave molecule composed of $D_s\bar{D}_{s0}$.
\begin{figure}[h!]
\begin{center}
\includegraphics[scale=0.45]{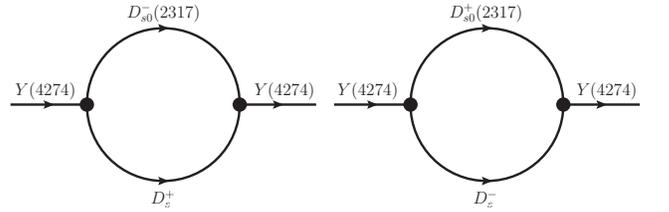}
\caption{Self-energy of the $Y(4274)$ state.} \label{isi}
\end{center}
\end{figure}

The coupling constant $g_{YD_s\bar{D}_{s0}}$ in Eq.~(\ref{eq1}) can be computed by the compositeness condition~\cite{Weinberg:1962hj,Salam:1962ap},
which indicates that the renormalization constants of a composite particle wave function should be zero, i.e.,
\begin{align}
Z_{Y}=1-\frac{d\Sigma^{T}_{Y}}{dk_0}|_{k_0=m_{Y}}=0\label{eqn3},
\end{align}
where the $\Sigma^{T}_{Y}$ is the transverse part of the mass operator and relates to its mass operator via the relation.
\begin{align}
\Sigma^{\mu\nu}_{Y}(p)=(g_{\mu\nu}-p^{\mu}p^{\nu}/p^2)\Sigma^{T}_{Y}+....
\end{align}
The concrete forms of the mass operators of the $Y(4274)$ corresponding to the diagrams in Fig.~\ref{isi} is
\begin{align}
\Sigma^{\mu\nu}_{Y}(k_0)&=\frac{g^2_{YD_s\bar{D}_{s0}}}{2}\int\frac{d^4k_1}{(2\pi)^4}\{\Phi^2[(k_1-k_0\omega_{D^{+}_s})^2]\frac{i}{k_1^2-m^2_{D^{+}_s}}\nonumber\\
                        &\times\frac{i}{(k_1-k_0)^2-m^2_{D^{-}_{s0}}}+\Phi^2[(k_1-k_0\omega_{D^{-}_s})^2]\frac{i}{k_1^2-m^2_{D^{-}_s}}\nonumber\\
                        &\times\frac{i}{(k_1-k_0)^2-m^2_{D^{+}_{s0}}}\}(k_0-2k_1)_{\mu}(k_0-2k_1)_{\nu}\label{eqn1},
\end{align}
where $k_0^2=m^2_{Y}$ with $k_0, m_{Y}$ denoting the four momenta and mass of the $Y(4274)$, respectively,  $k_1$, $m_{D_s}$, and $m_{D_{s0}}$ are
the four-momenta, the mass of the $D_s$ meson, and the mass of the $\bar{D}_{s0}$ meson, respectively.   With above preparations, we can obtain the coupling
constant of the $D_s\bar{D}_{s0}$ molecule to its components
\begin{align}
\frac{1}{g^2_{YD_s\bar{D}_{s0}}}&=\frac{m_Y}{8\pi^2{\cal{S}}}\int_{0}^{\infty}d\alpha\int_{0}^{\infty}d\beta\sum_{i=1}^{2}\nonumber\\
                          &\times\exp[-\frac{1}{\Lambda^2}(-{\cal{F}}_im^2_{Y}+{\cal{H}}_i+\frac{{\cal{C}}_i^2m_Y^2}{z})](\frac{{\cal{F}}_i}{z^3}-\frac{{\cal{C}}^2_i}{z^4})
\end{align}
where ${\cal{F}}_1=(2\omega^2_{D_s^{+}}+\beta),{\cal{F}}_2=(2\omega^2_{D_s^{-}}+\beta),{\cal{H}}_1=\alpha{}m^2_{D_s^{+}}+\beta{}m^2_{D^{-}_{s0}}$, ${\cal{H}}_2=\alpha{}m^2_{D_s^{-}}+\beta{}m^2_{D^{+}_{s0}}$, ${\cal{C}}_1=(2\omega_{D_s^{+}}+\beta)$,${\cal{C}}_2=(2\omega_{D_s^{-}}+\beta)$, $z=2+\alpha+\beta$, and ${\cal{S}}=1.0$ GeV.

\subsection{The decay $Y(4274)\to{}J/\psi{}\phi$}
\begin{figure}[htbp]
\begin{center}
\includegraphics[scale=0.45]{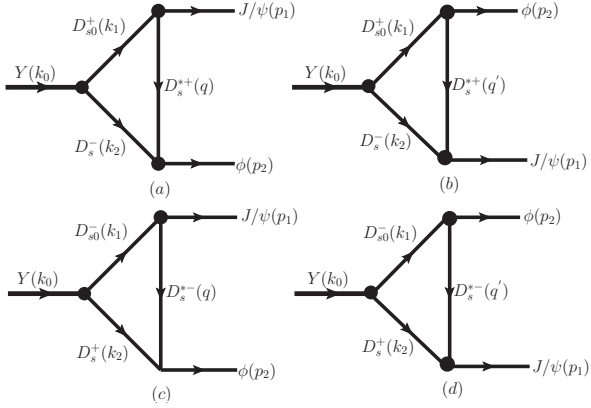}
\caption{Feynman diagrams for the $Y(4274)\to{}J/\psi\phi$ decay processes.
We also show the definitions of the kinematics($k_0,k_1,k_2,p_1,p_2,q$,and $q^{'}$)
used in the calculation.}\label{mku}
\end{center}
\end{figure}

Since the $Y(4274)$ was observed in the $J/\psi{}\phi$ invariant mass,  we first calculate the $J/\psi{}\phi$ two-body decay width of the $Y(4274)$
via the triangle diagrams show in Fig.~\ref{mku}.
The hadronic decay of the $D_s\bar{D}_{s0}$ molecular state into $J/\psi\phi$ mediated by the exchange of the $D_s^{*}$ meson.
To evaluate the diagrams, in addition to the Lagrangian in Eq.~\ref{eq1}, the following effective Lagrangians, responsible for vector meson
$V(=J/\psi,\phi)$ coupling to $D_s^{*}\bar{D}_{s0}$ are needed as well~\cite{Ma:2010xx}
\begin{align}
{\cal{L}}_{D_s^{*}\bar{D}_{s0}V}=g_{V\bar{D}_{s0}D_s^{*}}[D^{-}_{s0}D^{*+}_{s;\mu\nu}-D^{+}_{s0}D^{*-}_{s;\mu\nu}]V^{\mu\nu},
\end{align}
where $V^{\mu\nu}=\partial^{\mu}V^{\nu}-\partial^{\nu}V^{\mu}$.  The coupling constant $g_{J/\psi\bar{D}_{s0}D_s^{*}}=0.225$ GeV$^{-1}$
and $g_{\phi\bar{D}_{s0}D_s^{*}}=0.135$ GeV$^{-1}$ are got from Ref.~\cite{Ma:2010xx}.

To compute the $D_sD_s^{*}V$ vertices, we also need the following effective Lagrangian~\cite{Oset:2002sh,Bramon:1992kr}
\begin{align}
{\cal{L}}_{VVP}=\frac{G}{\sqrt{2}}\epsilon^{\mu\nu\alpha\beta}\langle\partial_{\mu}V_{\nu}\partial_{\alpha}V_{\beta}P\rangle\label{we1},
\end{align}
where $G=3h^2/(4\pi^2f)$ with $h=-G_Vm_{\rho}/(\sqrt{2}f^2)$, $f=0.093$ GeV, $G_V=0.069$ GeV~\cite{Oset:2002sh,Bramon:1992kr}, and $m_{\rho}=0.775$ GeV.
 $V_{\mu}$ and $P$ are standard $SU(4)$ matrices constructed with 16-plet of the vector meson contain $\rho$ and the
16-plet of pseudoscalar mesons containing the $\pi$, respectively.
\begin{equation}
V_{\mu}=
\left(
  \begin{array}{cccc}
    \frac{1}{\sqrt{2}}(\rho^{0}+\omega) & \rho^{+}                             &  K^{*+}     & \bar{D}^{*0} \\
    \rho^{-}                            & \frac{1}{\sqrt{2}}(-\rho^{0}+\omega) &  K^{*0}     & -D^{*-}       \\
     K^{*-}                             & \bar{K}^{*0}                         &  \phi       & D^{*-}_s     \\
     D^{*0}                             & -D^{*+}                               &  D^{*+}_{s} & J/\psi       \\
  \end{array}
\right)_{\mu},
\end{equation}
\begin{equation}
P=\sqrt{2}
\left(
  \begin{array}{cccc}
    \frac{\pi^{0}}{\sqrt{2}}+\frac{\eta}{\sqrt{6}}+\frac{\eta^{'}}{\sqrt{3}}   & \pi^{+}                  &  K^{+}     & \bar{D}^{0} \\
    \pi^{-}                   & -\frac{\pi^{0}}{\sqrt{2}}+\frac{\eta}{\sqrt{6}}+\frac{\eta^{'}}{\sqrt{3}} &  K^{0}     & -D^{-}       \\
     K^{-}                             & \bar{K}^{0}             &  -\sqrt{\frac{2}{3}}\eta+\frac{1}{\sqrt{3}}\eta^{'} & D^{-}_s     \\
     D^{0}                             & -D^{+}                                 &  D^{+}_{s}                            & \eta_c      \\
  \end{array}
\right).
\end{equation}

With above effective Lagrangians, we can get the decay amplitudes corresponding to the diagrams in Fig.~\ref{mku}
\begin{align}
{\cal{M}}_{a}&=\frac{G}{\sqrt{2}}g_{J/\psi{}\bar{D}_{s0}D^{*}_s}g_{YD_s\bar{D}_{s0}}\int\frac{d^4q}{(2\pi)^4}\Phi[(k_1\omega_{D^{-}_{s}}-k_2\omega_{D_{s0}^{+}})^2]\nonumber\\
                  &\times{}(p_{1\mu}g_{\nu\sigma}-p_{1\nu}g_{\mu\sigma})(q_{\mu}g_{\nu\eta}-q_{\nu}g_{\mu\eta})(g_{\eta\lambda}-\frac{q_{\eta}q_{\lambda}}{m^2_{D^{*+}_s}})\nonumber\\
                  &\times{}\epsilon^{\tau\lambda\alpha\beta}q_{\tau}p_{2\alpha}(k_1^{\varphi}-k_2^{\varphi})\epsilon^{Y}_{\varphi}(k_0)\epsilon^{*\sigma}_{J/\psi}(p_1)\epsilon^{*\phi}_{\beta}(p_2)\nonumber\\
                  &\times{}\frac{1}{q^2-m^2_{D^{*+}_{s}}}\frac{1}{k_2^2-m^2_{D^{-}_{s}}}\frac{1}{k_1^2-m^2_{D^{-}_{s0}}},\label{wq1}\\
{\cal{M}}_{b}&=\frac{G}{\sqrt{2}}g_{\phi\bar{D}_{s0}D^{*}_s}g_{YD_s\bar{D}_{s0}}\int\frac{d^4q^{'}}{(2\pi)^4}\Phi[(k_1\omega_{D^{-}_{s}}-k_2\omega_{D_{s0}^{+}})^2]\nonumber\\
             &\times{}(p_{2\lambda}g_{\sigma\tau}-p_{2\sigma}g_{\lambda\tau})(q^{'}_{\lambda}g_{\sigma\eta}-q^{'}_{\sigma}g_{\lambda\eta})(g_{\nu\eta}-\frac{q^{'}_{\nu}q^{'}_{\eta}}{m^2_{D^{*+}_s}})\nonumber\\
             &\times{}\epsilon^{\mu\nu\alpha\beta}q^{'}_{\mu}p_{1\alpha}(k_1^{\varphi}-k_2^{\varphi})\epsilon^{Y}_{\varphi}(k_0)\epsilon^{*\phi}_{\tau}(p_2)\epsilon^{*J/\psi}_{\beta}(p_1)\nonumber\\
                  &\times{}\frac{1}{q^{'2}-m^2_{D^{*+}_{s}}}\frac{1}{k_2^2-m^2_{D^{-}_{s}}}\frac{1}{k_1^2-m^2_{D^{-}_{s0}}},\label{wq2}\\
{\cal{M}}_{c}&={\cal{M}}_{a}(D^{+}_{s0}\to{}D^{-}_{s0},D^{-}_{s}\to{}D^{+}_{s},D^{*+}_s\to{}D^{*-}_s),\label{wq3}\\
{\cal{M}}_{d}&={\cal{M}}_{b}(D^{+}_{s0}\to{}D^{-}_{s0},D^{-}_{s}\to{}D^{+}_{s},D^{*+}_s\to{}D^{*-}_s).\label{wq4}
\end{align}

\subsection{The decay $Y(4274)\to{}\chi_{c0}\eta$ and $\chi_{c1}\eta$}
In this section, we compute the other possible decays of $Y(4274)$ with $D_s\bar{D}_{s0}$ molecular.
Fig.~\ref{mku-other} shows the hadronic decay of the $D_s\bar{D}_{s0}$ molecular state into
$\chi_{c0}\eta$ and $\chi_{c1}\eta$ mediated by the exchange of $D_s$ meson.
\begin{figure}[htbp]
\begin{center}
\includegraphics[scale=0.45]{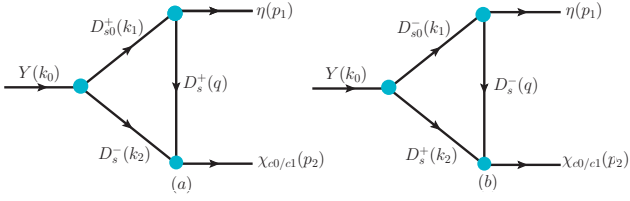}
\caption{Diagrams for $Y(4274)\to{}\eta\chi_{c0/c1}$ decay processes.}\label{mku-other}
\end{center}
\end{figure}
The ingredients need are the $\chi_{c0}D^{+}_sD^{-}_s$ and $\chi_{c1}D^{+}_sD^{-}_s$ Lagrangians~\cite{Colangelo:2003sa,Casalbuoni:1996pg}
\begin{align}
&{\cal{L}}_{\chi_{c1}D^{+}_sD^{-}_s}=-ig_{\chi_{c1}D^{+}_sD^{-}_s}(D_s^{+}\partial_{\mu}D_s^{-}-\partial_{\mu}D_s^{+}D_s^{-})\chi_{c1}^{\mu},\label{yut}\\
&{\cal{L}}_{\chi_{c0}D_s^{+}D_s^{-}}=-g_{\chi_{c0}D_s^{+}D_s^{-}}\chi_{c0}D_s^{+}D_s^{-}.\label{yuq}
\end{align}
The coupling constants $g_{\chi_{c1}D^{+}_sD^{-}_s}$ and $g_{\chi_{c0}D_s^{+}D_s^{-}}$ can be fixed from the heavy
quark field theory~\cite{Casalbuoni:1996pg}.

Moreover, the effective Lagrangian, responsible for the coupling of $D_{s0}$ to $\eta{}D_s$, is needed as well~\cite{Huang:2019qmw}
\begin{align}
{\cal{L}}_{D_{s0}D_s\eta}=g_{D_{s0}D_s\eta}D_{s0}D_s\eta,
\end{align}
where the coupling constant is found to be $g_{D_{s0}D_s\eta}=6.40$~\cite{Gamermann:2006nm}.

Thus, we can obtain the following amplitudes for the decays $Y(4274)\to{}\chi_{c0}\eta$
\begin{align}
{\cal{M}}_{a}&=\frac{g_{\eta\bar{D}_{s0}D_s}g_{YD_s\bar{D}_{s0}}g_{D_s^{-}D_s^{+}\chi_{c0}}}{\sqrt{2}}\int\frac{d^4q}{(2\pi)^4}\Phi[(k_1\omega_{D^{-}_{s}}-k_2\omega_{D_{s0}^{+}})^2]\nonumber\\
             &\times{}(k_1^{\varphi}-k_2^{\varphi})\epsilon^{Y}_{\varphi}(k_0)\frac{1}{q^2-m^2_{D_s^{+}}}\frac{1}{k_2^2-m^2_{D^{-}_{s}}}\frac{1}{k_1^2-m^2_{D^{+}_{s0}}},\label{swq1}\\
{\cal{M}}_{b}&=-\frac{g_{\eta\bar{D}_{s0}D_s}g_{YD_s\bar{D}_{s0}}g_{D_s^{-}D_s^{+}\chi_{c0}}}{\sqrt{2}}\int\frac{d^4q}{(2\pi)^4}\Phi[(k_1\omega_{D^{+}_{s}}-k_2\omega_{D_{s0}^{-}})^2]\nonumber\\
             &\times{}(k_1^{\varphi}-k_2^{\varphi})\epsilon^{Y}_{\varphi}(k_0)\frac{1}{q^2-m^2_{D_s^{-}}}\frac{1}{k_2^2-m^2_{D^{+}_{s}}}\frac{1}{k_1^2-m^2_{D^{-}_{s0}}}\label{swq3},
\end{align}
and $Y(4274)\to{}\chi_{c1}\eta$
\begin{align}
{\cal{M}}_{a}&=\frac{g_{\eta\bar{D}_{s0}D_s}g_{YD_s\bar{D}_{s0}}g_{D_s^{-}D_s^{+}\chi_{c1}}}{\sqrt{2}}\int\frac{d^4q}{(2\pi)^4}\Phi[(k_1\omega_{D^{-}_{s}}-k_2\omega_{D_{s0}^{+}})^2]\nonumber\\
             &\times{}(k_1^{\varphi}-k_2^{\varphi})\epsilon^{Y}_{\varphi}(k_0)(k_2^{\mu}-q^{\mu})\epsilon^{\chi_{c1}}_{\mu}(p_2)\nonumber\\
             &\times\frac{1}{q^2-m^2_{D_s^{+}}}\frac{1}{k_2^2-m^2_{D^{-}_{s}}}\frac{1}{k_1^2-m^2_{D^{+}_{s0}}},\label{swq11}\\
{\cal{M}}_{b}&=-\frac{g_{\eta\bar{D}_{s0}D_s}g_{YD_s\bar{D}_{s0}}g_{D_s^{-}D_s^{+}\chi_{c1}}}{\sqrt{2}}\int\frac{d^4q}{(2\pi)^4}\Phi[(k_1\omega_{D^{+}_{s}}-k_2\omega_{D_{s0}^{-}})^2]\nonumber\\
             &\times{}(k_1^{\varphi}-k_2^{\varphi})\epsilon^{Y}_{\varphi}(k_0)(k_2^{\mu}-q^{\mu})\epsilon^{\chi_{c1}}_{\mu}(p_2)\nonumber\\
             &\times\frac{1}{q^2-m^2_{D_s^{-}}}\frac{1}{k_2^2-m^2_{D^{+}_{s}}}\frac{1}{k_1^2-m^2_{D^{-}_{s0}}}\label{swq31}.
\end{align}
The minus sign in Eqs.~(\ref{swq3}) and (\ref{swq31}) come from the flavor function that shown in Eq.~(\ref{eqr3}).
Then, we find ${\cal{M}}[Y(4274)\to{}\chi_{c0/c1}\eta]_{Total}={\cal{M}}_{a}+{\cal{M}}_{b}=0$.

\subsection{The decay $Y(4274)\to{}D^{*}_s\bar{D}_s$}
Fig.~\ref{mku-other} shows the hadronic decay of the $D_s\bar{D}_{s0}$ molecular state into
 $D^{*}_s\bar{D}_s$ mediated by the exchange of $\eta$ meson.
\begin{figure}[htbp]
\begin{center}
\includegraphics[scale=0.45]{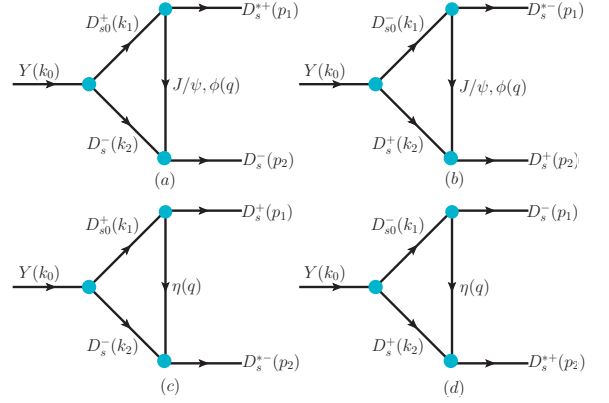}
\caption{Diagrams for $Y(4274)$ decay processes:(a,d)
$Y(4274)\to{}D^{*+}_sD^{-}_s$, and (b,c)
$Y(4274)\to{}D^{*-}_sD^{+}_s$.}\label{mku-other}
\end{center}
\end{figure}
The ingredients need are the vector($V$)-pseudoscalar($P$)-pseudoscalar($P$) Lagrangian
\begin{align}
{\cal{L}}_{VPP}=-ig\langle[P,\partial_{\mu}P]V^{\mu}\rangle,\label{yut}
\end{align}
The coupling $g$ is fixed from the strong decay width of $K^{*}\to{}K\pi$.  With the help of Eq.~(\ref{yut}), the two-body decay width
 $\Gamma(K^{*+}\to{}K^{0}\pi^{+})$ is related to $g$ as
 \begin{align}
 \Gamma(K^{*+}\to{}K^{0}\pi^{+})=\frac{g^2}{6\pi{}m^2_{K^{*+}}}{\cal{P}}^3_{\pi{}K^{*}}=\frac{2}{3}\Gamma_{K^{*+}},
 \end{align}
where ${\cal{P}}_{\pi{}K^{*}}$ is the three-momentum of the $\pi$ in the rest frame of the  $K^{*}$.
Using the experimental strong decay width ($\Gamma_{K^{*+}}=50.3\pm{}0.8$ MeV) and the masses of the particles needed in
the present work~\cite{Zyla:2020zbs} we obtain $g=4.61$.

Thus, we can obtain the following amplitudes for the decay $Y(4274)\to{}D^{*}_s\bar{D}_s$
\begin{align}
{\cal{M}}_{a}&=\frac{-gg_{V\bar{D}_{s0}D_s^{*}}g_{YD_s\bar{D}_{s0}}}{\sqrt{2}}\int\frac{d^4q}{(2\pi)^4}\Phi[(k_1\omega_{D^{-}_{s}}-k_2\omega_{D_{s0}^{+}})^2]\nonumber\\
             &\times{}(p_{1\mu}g_{\nu\eta}-p_{1\nu}g_{\mu\eta})(q_{\mu}g_{\nu\lambda}-q_{\nu}g_{\mu\lambda})(g_{\lambda\alpha}-\frac{q_{\lambda}q_{\alpha}}{m^2_{V}})\nonumber\\
             &\times{}(k_{2\alpha}+p_{2\alpha})(k_{1\omega}-k_{2\omega})\epsilon^{*\eta}_{D^{*+}_{s}}(p_1)\epsilon^{\omega}_{Y}(k_0)\nonumber\\
             &\times{}\frac{1}{q^{2}-m^2_{V}}\frac{1}{k_2^2-m^2_{D^{-}_{s}}}\frac{1}{k_1^2-m^2_{D^{+}_{s0}}},\label{swq3}\\
{\cal{M}}_{b}&={\cal{M}}_{b}(D^{+}_{s0}\to{}D^{-}_{s0},D^{-}_{s}\to{}D^{+}_{s},D^{*+}_s\to{}D^{*-}_s),\label{wqd}\\
{\cal{M}}_{c}&={\cal{M}}_{f}(D^{-}_{s0}\to{}D^{+}_{s0},D^{+}_{s}(k_2)\to{}D^{-}_{s}(k_2),D^{*+}_s\to{}D^{*-}_s)\\
{\cal{M}}_{d}&=-\frac{gg_{YD_s\bar{D}_{s0}}g_{D_{s0}D_s\eta}}{\sqrt{6}}\int\frac{d^4q}{(2\pi)^4}\Phi[(k_1\omega_{D^{+}_{s}}-k_2\omega_{D_{s0}^{-}})^2]\nonumber\\
             &\times{}(q_{\mu}-k_{2\mu})(k_{1\nu}-k_{2\nu})\epsilon^{Y}_{\nu}(k_0)\epsilon^{*\mu}_{D^{*+}_{s}}(p_1)\nonumber\\
             &\times\frac{1}{q^{2}-m^2_{\eta}}\frac{1}{k_2^2-m^2_{D^{+}_{s}}}\frac{1}{k_1^2-m^2_{D^{-}_{s0}}},\label{swqqw}
\end{align}

\subsection{The decay $Y(4274)\to{}D_s\bar{D}_s\pi^0$}
\begin{figure}[htbp]
\begin{center}
\includegraphics[bb=70 570 600 710, clip, scale=0.5]{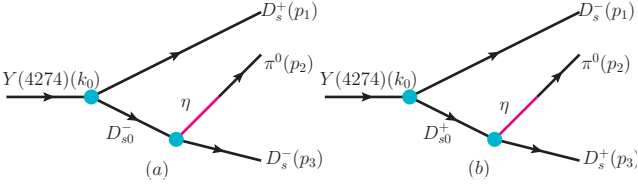}
\caption{Feynman diagrams for $Y(4274)\to{}D_s\bar{D}_s\pi^0$ decay processes.
We also show the definitions of the kinematics($k_0,p_1,p_2,p_3$)
used in the calculation.}\label{mku-santi}
\end{center}
\end{figure}

Now we turn to $D_s\bar{D}_s\pi^0$ three-body decay channel of $Y(4274)$.  Under the $D_s\bar{D}_{s0}$ molecular state assignment,
$Y(4274)$ first dissociates into $D_s^{+}D_{s0}^{-}$ or $D_s^{-}D_{s0}^{+}$.  Then, the decay $Y(4274)\to{}D_s^{+}D_s^{-}\pi^0$
occur via the transitions $D_{s0}^{\pm}\to{}D^{\pm}_{s}\pi^0$, where $D_{s0}^{\pm}$ decay into $D^{\pm}_{s}\pi^0$ by considering
the $\eta-\pi^0$ mixing mechanism~\cite{BaBar:2003oey,Liu:2006jx}.  The relevant Feynman diagrams ars shown in Fig.~\ref{mku-santi}.

The Lagrangian including $\eta-\pi^0$ mixing for $\pi^0D_{s0}D_s$ have been constructed in Ref.~\cite{He:2011ed} and
in the form
\begin{align}
{\cal{L}}_{\pi^0D_{s0}D_s}&=g_{\pi^0D_{s0}D_s}\pi^0D_{s0}D_s,
\end{align}
where the coupling constant $g_{\pi^0D_{s0}D_s}$ can be extracted by the relation
\begin{align}
\Gamma(D_{s0}^{\pm}\to{}D_{s}^{\pm}\pi^0)=g^2_{\pi^0D_{s0}D_s}\frac{|\vec{{\cal{P}}}_{\pi^0}|}{8\pi{}m^2_{D^{\pm}_{s0}}}.
\end{align}
In above, ${\cal{P}}_{\pi^0}$ is the three-momentum of $\pi$ in the rest frame of $D_{s0}$.

At present, the experimental partial decay width on $D_{s0}^{\pm}\to{}D_{s}^{\pm}\pi^0$ is absent.  We only find the absolute branching fraction
$B(D_{s0}^{\pm}\to{}D_{s}^{\pm}\pi^0)$ is measured as $1.00^{+0.00}_{-0.14}$(stat)$^{+0.00}_{-0.14}$(syst)~\cite{BESIII:2017vdm}
and the upper limit on $D_{s0}$ width is 3.8 MeV at the 95$\%$ confidence level (C.L.)~\cite{BaBar:2006eep}.  In this work, we focus
on whether $Y(4274)$ can be a $P$- wave $D_{s}\bar{D}_{s0}$ molecular state and hope that the theoretical maximum decay width cannot
be compared with the experimental data.  Thus, we use upper limit $\Gamma{(D_{s0})}=3.8$ MeV to determine the
coupling constant $g_{\pi^0D_{s0}D_s}=1.3124$ GeV and in further calculation.

The general expression of the decay amplitude $Y(4274)\to{}D_s\bar{D}_s\pi^0$ is
\begin{align}
{\cal{M}}&_{a/b}=\frac{g_{\pi^0D_{s0}D_s}g_{YD_s\bar{D}_{s0}}}{\sqrt{2}}(q_{\mu}-p_{1\mu})\epsilon^{\mu}_{Y}(k_0)\nonumber\\
         &\times\Phi[(p_1\omega_{\bar{D}_{s0}}-q\omega_{D_s})^2]\frac{1}{q^2-m^2_{D_{s0}}+im_{D_{s0}}\Gamma_{D_{s0}}},
\end{align}
 where $q=p_2+p_3$.

\subsection{The decay $Y(4274)\to{}D\bar{D}^{*}$, $K\bar{K}^{*}$, and $\phi\phi$}
\begin{figure}[htbp]
\begin{center}
\includegraphics[bb=50 555 600 710, clip, scale=0.45]{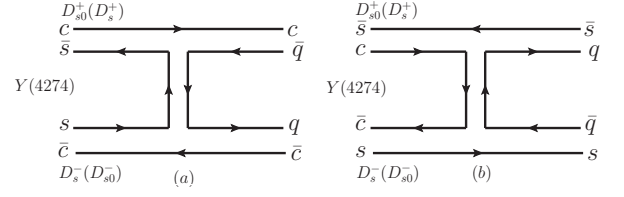}
\caption{The $Y(4274)$ decay process via the OZI mechanism.
$q=u,d$ quarks for diagram (a) and $q=u,d,s$ quarks for diagram (b).}\label{mku-other-ozi}
\end{center}
\end{figure}

\begin{figure}[htbp]
\begin{center}
\includegraphics[scale=0.45]{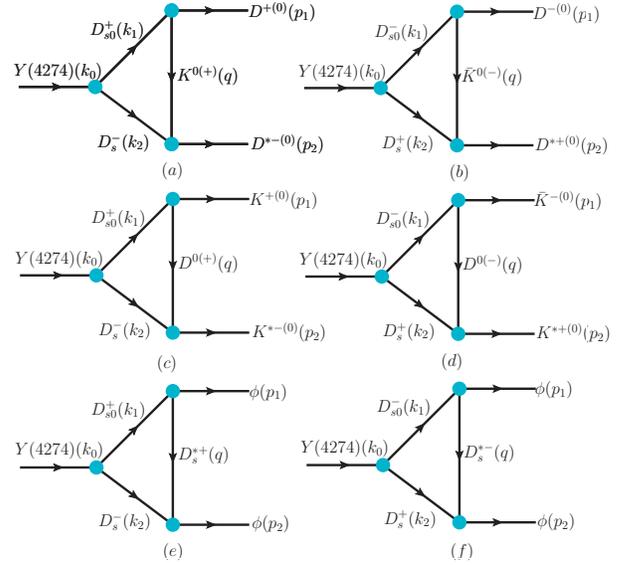}
\caption{Diagrams for $Y(4274)$ decay processes:(a,b) $Y(4274)\to{}D\bar{D}^{*}$,(c,d)
$Y(4274)\to{}K\bar{K}^{*}$, and (e,f) $Y(4274)\to{}\phi\phi$.}\label{mky-other-ozi}
\end{center}
\end{figure}

In this section, we calculate the two-body Okubo-Zweig-Iizuka (OZI) allowed strong decays $Y(4274)\to{}D\bar{D}^{*}$,
$K\bar{K}^{*}$, and $\phi\phi$.  The processes is described as a quark-antiquark pair $c\bar{c}$ or $s\bar{s}$
annihilation in the initial state.  Meanwhile,  a light quark-antiquark pair is created and then regroups into two
outgoing hadrons by a quark rearrangement process.  The decay of $Y(4274)$ into other channels, such as $Y(4274)\to{}\pi\bar{\pi}$,
are ignored because in these processes, which involve the creation or annihilation of two $\bar{q}q$ $(q=u,d,s,c)$
quark pairs, are usually strongly suppressed.  The quark-level diagrams are depicted in Fig.~\ref{mku-other-ozi}.
And corresponding hadron-level diagrams are in Fig.~\ref{mky-other-ozi}.

Besides the Lagrangians above, the effective Lagrangian of $KDD_{s0}$ vertex is also needed~\cite{Huang:2019qmw}
\begin{align}
{\cal{L}}_{KDD_{s0}}&=g_{KDD_{s0}}KDD_{s0},
\end{align}
where the coupling constant $g_{KDD_{s0}}=\sqrt{2}g_{K^{-}D^0D^{-}_{s0}}=10.21$ is obtained from the coupling of the $D_{s0}$ to the $DK$ channel
in isospin $I=0$~\cite{Gamermann:2006nm}.   Putting all pieces together, we obtain the amplitudes for $Y(4274)\to{}D\bar{D}^{*}$, $K\bar{K}^{*}$, and $\phi\phi$
which correspond to the diagrams in Fig.~\ref{mky-other-ozi},
\begin{align}
{\cal{M}}_{a}&=\frac{g_{KDD_{s0}}gg_{Y}}{\sqrt{2}}\int\frac{d^4q}{(2\pi)^4}\Phi[(k_1\omega_{D^{-}_{s}}-k_2\omega_{D_{s0}^{+}})^2]\nonumber\\
            &\times{}(q_{\mu}-k_{2\mu})(k_{1\nu}-k_{2\nu})\epsilon_{Y}^{\nu}(k_0)\epsilon^{*\mu}_{D^{*-(0)}}(p_2)\nonumber\\
             &\times\frac{1}{q^{2}-m^2_{K^{0(+)}}}\frac{1}{k_2^2-m^2_{D^{-}_{s}}}\frac{1}{k_1^2-m^2_{D^{+}_{s0}}},
            \end{align}
\begin{align}
{\cal{M}}_{b}&=-\frac{g_{KDD_{s0}}gg_{Y}}{\sqrt{2}}\int\frac{d^4q}{(2\pi)^4}\Phi[(k_1\omega_{D^{+}_{s}}-k_2\omega_{D_{s0}^{-}})^2]\nonumber\\
              &\times{}(q_{\mu}-k_{2\mu})(k_{1\nu}-k_{2\nu})\epsilon_{Y}^{\nu}(k_0)\epsilon^{*\mu}_{D^{*+(0)}}(p_2)\nonumber\\
              &\times\frac{1}{q^{2}-m^2_{\bar{K}^{0(-)}}}\frac{1}{k_2^2-m^2_{D^{+}_{s}}}\frac{1}{k_1^2-m^2_{D^{-}_{s0}}},\\
{\cal{M}}_{c}&={\cal{J}}\frac{g_{KDD_{s0}}gg_{Y}}{\sqrt{2}}\int\frac{d^4q}{(2\pi)^4}\Phi[(k_1\omega_{D^{-}_{s}}-k_2\omega_{D_{s0}^{+}})^2]\nonumber\\
             &\times{}(q_{\mu}-k_{2\mu})(k_{1\nu}-k_{2\nu})\epsilon_{Y}^{\nu}(k_0)\epsilon^{*\mu}_{K^{*-(0)}}(p_2)\nonumber\\
             &\times\frac{1}{q^{2}-m^2_{D^{0(+)}}}\frac{1}{k_2^2-m^2_{D^{-}_{s}}}\frac{1}{k_1^2-m^2_{D^{+}_{s0}}},\\
{\cal{M}}_{d}&={\cal{J}}\frac{g_{KDD_{s0}}gg_{Y}}{\sqrt{2}}\int\frac{d^4q}{(2\pi)^4}\Phi[(k_1\omega_{D^{+}_{s}}-k_2\omega_{D_{s0}^{-}})^2]\nonumber\\
             &\times{}(q_{\mu}-k_{2\mu})(k_{1\nu}-k_{2\nu})\epsilon_{Y}^{\nu}(k_0)\epsilon^{*\mu}_{K^{*+(0)}}(p_2)\nonumber\\
             &\times\frac{1}{q^{2}-m^2_{D^{0(-)}}}\frac{1}{k_2^2-m^2_{D^{+}_{s}}}\frac{1}{k_1^2-m^2_{D^{-}_{s0}}},\\
{\cal{M}}_{e}&=-\frac{Gg_{Y}g_{\phi\bar{D}_{s0}D_s^{*}}}{2}\int\frac{d^4q}{(2\pi)^4}\Phi[(k_1\omega_{D^{+}_{s}}-k_2\omega_{D_{s0}^{-}})^2]\nonumber\\
             &\times{}(q^{\mu}g^{\nu\eta}-q^{\nu}g^{\mu\eta})(p_{1\mu}g_{\nu\rho}-p_{1\nu}g_{\mu\rho})(-g^{\eta\lambda}+q^{\eta}q^{\lambda}/m^2_{D_s^{*+}})\nonumber\\
             &\times{}\epsilon^{\tau\lambda\alpha\beta}q_{\tau}p_{2\alpha}(k_{1\sigma}-k_{2\sigma})\epsilon_{\phi}^{*\rho}(p_1)\epsilon_{\phi}^{*\beta}(p_2)\epsilon_{Y}^{\sigma}(k_0)\nonumber\\
             &\times{}\frac{1}{q^2-m^2_{D_s^{*+}}}\frac{1}{k_2^2-m^2_{D^{-}_{s}}}\frac{1}{k_1^2-m^2_{D^{+}_{s0}}},\\
{\cal{M}}_{f}&={\cal{M}}_{e}(D_{s0}^{+}\to{}D_{s0}^{-},D_s^{-}\to{}D_s^{+},D_s^{*+}\to{}D_{s}^{*-}),
\end{align}
where ${\cal{J}}$=1 and $-1$ are for $D^{0}(D^0)$ and $D^{+}(D^{-})$ exchange, respectively.

Once the amplitudes are determined, the corresponding partial decay widths can be obtained, which reads,
\begin{align}
\Gamma(Y(4274)\to MB)&=\frac{1}{24\pi}\frac{|\vec{p}_1|}{m^2_{Y}}\overline{|{\cal{M}}|^2},\\
\Gamma(Y(4274)\to D_s\bar{D}_s\pi^0)&=\frac{1}{3(2\pi)^5}\frac{1}{16m^2_{Y}}\overline{|{\cal{M}}|^2}|\vec{p}^{*}_3||\vec{p}_1|\nonumber\\
                                    &\times{}dm_{\pi^0D_{s}^{\pm}}d\Omega^{*}_{p_3}d\Omega_{p_1}\label{eq31},
\end{align}
where $J$ is the total angular momentum of the Y(4274) state, the $|\vec{p}_1|$ is the three-momenta of the decay products
in the center of mass frame,  the overline indicates the sum over the polarization vectors of the final hadrons, and $MB$
denotes the decay channel of $MB$, i.e., $J/\psi{}\phi$, $\chi_{c0}\eta$, $\chi_{c1}\eta$,
$D^{*}_s\bar{D}_s$, $D\bar{D}^{*}$, $K\bar{K}^{*}$ and $\phi\phi$.   In Eq.~\ref{eq31}, the
$\vec{p}^{*}_3$ and $\Omega^{*}_{p_3}$ are the momentum and angle of the particle $D^{\pm}_{s}$ in the rest frame of $D^{\pm}_{s}$ and $\pi^0$, respectively,
and $\Omega_{p_1}$ is the angle of $D_s^{\pm}$ in the rest frame of the decaying particle $Y(4274)$. $m_{\pi^0D_{s}^{\pm}}$
is the invariant mass for $D^{\pm}_{s}$ and $\pi^0$ and must meet $m_{D^{\pm}_{s0}}+m_{\pi^0}\leq{}m_{\pi^0D_{s}^{\pm}}\leq{}m_Y-m_{D^{\pm}_{s}}$.

\section{RESULTS and Discussions}
In this work, we study the strong decays of the $Y(4274)$ to the two-body final states $J/\psi\phi, \chi_{c0}\eta, \chi_{c1}\eta,
D^{*}_s\bar{D}_s$, $D\bar{D}^{*}$, $K\bar{K}^{*}$, $\phi\phi$ and three-body decay into $\pi^0D_s\bar{D}_s$ assuming that $Y(4274)$
is a $D_s\bar{D}_{s0}(2317)$ molecular state.  In order to obtain the decay width shown in Figs.~\ref{mku}, \ref{mku-other},
and \ref{mku-santi}, the coupling constant $g_{YD_s\bar{D}_{s0}}$ should be computed first.

\begin{figure}[htbp]
\begin{center}
\includegraphics[bb=-20 140 750 550, clip, scale=0.35]{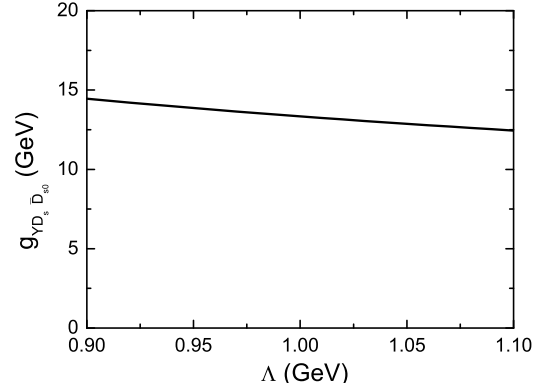}
\caption{The $\Lambda$ dependence of the coupling $g_{YD_s\bar{D}_{s0}}$
estimated from compositeness condition.}\label{couplin-constant}
\end{center}
\end{figure}
According to the compositeness condition that we introduced in Eq.~\ref{eqn3}, $\Lambda$ dependence of the coupling constant
$g_{YD_s\bar{D}_{s0}}$ is computed.   With a value of cutoff $\Lambda=0.9-1.1$ GeV, the corresponding coupling constants shown in
Fig~\ref{couplin-constant}.  We note that they decrease slowly with the increase of cut-off, and the coupling constant is almost
independent of $\Lambda$,  where $Y(4274)$ is $P-$wave $D_s\bar{D}_{s0}$ molecular state.  According to the studies
in Refs~\cite{Faessler:2007gv,Faessler:2007us,Dong:2008gb,Dong:2009uf,Dong:2009yp,Dong:2017rmg,Dong:2014ksa,Dong:2014zka,Dong:2013kta,
Dong:2013iqa,Dong:2013rsa,Dong:2012hc,Dong:2011ys,Dong:2010xv,Dong:2010gu,Dong:2009tg,Dong:2017gaw,Yang:2021pio,Zhu:2020jke},
a typical value of $\Lambda=1.0$ GeV is often employed.  Thus, in this work we take $\Lambda$ = 1.0 GeV and the corresponding
coupling constants are $g_{YD_s\bar{D}_{s0}}=13.34^{+1.11}_{-0.89}$ GeV, which the error reflects variation in $\Lambda$
from $0.9$ to $1.1$ GeV.
\begin{figure}[htbp]
\begin{center}
\includegraphics[bb=20 140 600 560, clip,scale=0.40]{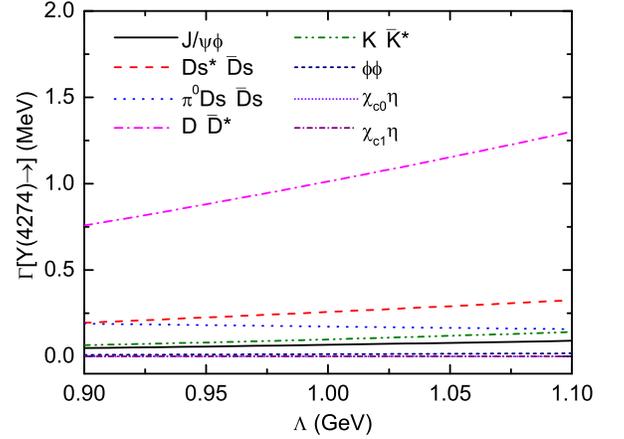}
\caption{(Color line) Partial decay widths of the $Y(4274)\to{}J/\psi{}\phi$ (black solid line),
$Y(4274)\to{}\pi^0D_s\bar{D}_s$ (blue dot line), $Y(4274)\to{}D_s^{*}\bar{D}_s$ (red dash line),
$Y(4274)\to{}\chi_{c0}\eta$ (violet short dot line), $Y(4274)\to{}D\bar{D}^{*}$ (magenta dash dot line),
$Y(4274)\to{}K\bar{K}^{*}$ (olive dash dot dot line), $Y(4274)\to{}\chi_{c1}\eta$ (purple short dash dot line),
and $Y(4274)\to{}\phi\phi$ (navy short dash line).}\label{width}
\end{center}
\end{figure}

Once the coupling constant $g_{YD_s\bar{D}_{s0}}=13.34^{+1.11}_{-0.89}$ GeV are determined, the decay widths of
$Y(4274)$ can be calculated straightforwardly.  In Fig.~\ref{width}, we show the partial decay widths of
$Y(4274)\to{}J/\psi\phi, \chi_{c0}\eta, \chi_{c1}\eta ,D^{*}_s\bar{D}_s$, $\pi^0D_s\bar{D}_s$, $D\bar{D}^{*}$, $K\bar{K}^{*}$,
and $\phi\phi$ as a function of cutoff parameter $\Lambda$.   We find that the estimated two-body decay widths increases
with increase of cut-off and are all insensitive to cut-off parameter $\Lambda$.  While the $Y(4274)\to{}\pi^0D_s\bar{D}_s$
three-body decay decreases, but very slowly.  We also find that the partial decay width is
the largest for transition $Y(4274)\to{}D\bar{D}^{*}$.  Thanks the flavor symmetry of the wave function that shown in Eq.~(\ref{eqr3}),
the $Y(4274)\to{}\chi_{c0}\eta$ and $\chi_{c1}\eta$ two-body decay widths are of the order of about 0.0 MeV.
We also note that the three-body transition strength is quite small, and the decay width is of the order of 0.158-0.190 MeV.
The small three-body decay width can be easy understand due to the decay $Y(4274)\to D_s\bar{D}_s\pi^0$ is an isospin violation
process.

Two reasons can help us to understand why transition $Y(4274)\to{}D\bar{D}^{*}$ provides the dominant contribution.
First is that transition $Y(4274)\to{}D\bar{D}^{*}$ is $s$-wave decay, and the lowest angular momentum gives the dominant
contribution.  Larger $Y(4274)\to{}D\bar{D}^{*}$ decay can also be understood due to the main component of $D_{s0}$
is $DK$, and the coupling constant related to this vertex is larger than the others.  The same $D_{s0}DK$ coupling also
exist in the $Y(4274)\to{}K\bar{K}^{*}$ reaction.  However, its partial decay width is small.  A possible explanation
is that a light quark-antiquark pair creation or annihilation in Fig.~\ref{mku-other-ozi} is easier than that of a heavy
quark-antiquark pair.
\begin{figure}[htbp]
\begin{center}
\includegraphics[bb=20 140 600 560, clip,scale=0.40]{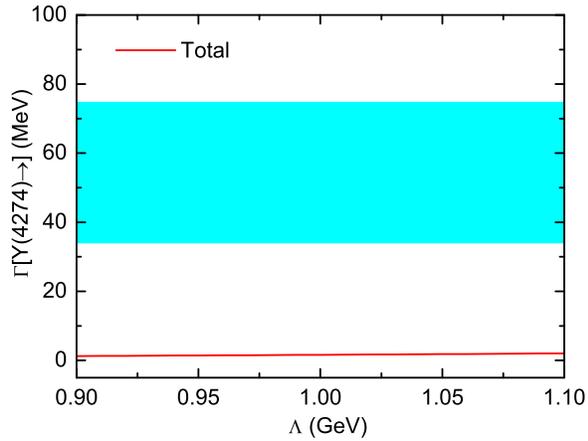}
\caption{(Color line) The total decay width of the $Y(4274)$.
The cyan bands denote the experimental total width~\cite{LHCb:2016axx,LHCb:2016nsl}.}\label{total-width}
\end{center}
\end{figure}

We also show the dependence of the total decay width on cutoff $\Lambda$ in Fig.~\ref{total-width}.   In present calculation,
we vary $\Lambda$ from 0.9 to 1.1 GeV.  In this $\Lambda$ range, the total decay width increases, and predicted decay width
$\Gamma_{Y}$=1.25-2.0 MeV is much smaller than the experimental width, which disfavors $Y(4274)$ in a $D_s\bar{D}_{s0}$
molecular picture.  If we increase $\Lambda$ to higher values,  the total widths of $Y(4274)$ cannot be reproduced until a much
larger $\Lambda$ value of about 9.6 is adopted.   Unfortunately, there are no such studies on taking $\Lambda=9.6$ or higher,
which is reasonable.  Hence, the assignment as a $P-$wave $D_s\bar{D}_{s0}$ molecular state is impossible for $Y(4274)$ based
on the total decay width experimentally measured.   This is quite different from the conclusion in Ref.~\cite{He:2016pfa}
that the interaction between a $D_s$ meson and a $\bar{D}_{s0}$ meson is strong enough to form a bound state with a mass about
4274 MeV, which can be associated to $Y(4274)$.  Comparing our results with those in Ref.~\cite{He:2016pfa}, it seems that a
study of the spectroscopy alone does not give a complete picture of its nature.

Combination our results in Fig.~\ref{total-width} with conclusion that in Ref.~\cite{He:2016pfa}, we predict a $P-$wave $D_s\bar{D}_{s0}$
molecular that we marked as $Y^{'}(4274)$ may exist.  Taking a typical value $\Gamma(D^{\pm}_{s0}\to\pi{}D^{\pm}_s)=79.3$ KeV~\cite{Faessler:2007gv},
where $D_{s0}$ is assumed to be a $DK$ bound state, the corresponding partial decay widths are $\Gamma[Y^{'}(4274)\to{}D_s^{*}\bar{D}_s]=0.20-0.33$ MeV, $\Gamma[Y^{'}(4274)\to{}J/\psi\phi]=0.048-0.090$, $\Gamma[Y^{'}(4274)\to{}\pi^0{}D_s\bar{D}_s]=0.0066-0.0080$ MeV,
$\Gamma[Y^{'}(4274)\to{}D\bar{D}^{*}]=0.76-1.30$ MeV, $\Gamma[Y^{'}(4274)\to{}K\bar{K}^{*}]=0.065-0.14$ MeV, $\Gamma[Y^{'}(4274)\to{}\chi_{c0}\eta]=0.0$ MeV, $\Gamma[Y^{'}(4274)\to{}\chi_{c1}\eta]=0.0$ MeV, and $\Gamma[Y^{'}(4274)\to{}\phi\phi]=0.0089-0.018$ MeV,
which yields a total decay width of 1.06-1.84 MeV.  And find that the transition $Y^{'}(4274)\to{}D\bar{D}^{*}$ is the main decay channel,
almost saturating the total width.  The experimental measurements for this strong decay process could be a crucial to observe such a state.

\section{summary}
In this work, inspired by the studies in Ref.~\cite{He:2016pfa} that showed the likely existence of a $D_s\bar{D}_{s0}$ bound state, we have
studied its partial decay widths into $Y(4274)\to{}J/\psi\phi$, $\chi_{c0}\eta$, $\chi_{c1}\eta$, $\phi\phi$, $\eta\eta$, $D^{*}_s\bar{D}_s$, 
$D\bar{D}^{*}$, $K\bar{K}^{*}$ and $\pi^0\bar{D}_sD_s$.  These decays involve the treatment of $Y(4274)$ state as a quasi-bound state of 
$D_s\bar{D}_{s0}$ and utilizing the Weinberg compositeness condition to determine the corresponding coupling.  Our studies find the $P-$wave 
$D_s\bar{D}_{s0}$ assignments for $Y(4274)$ is not allowed, it may be a compact tetraquark state or diquark-antidiquark state~\cite{Chen:2016oma,Wang:2016dcb,Stancu:2009ka,Agaev:2017foq,Lu:2016cwr,
Gui:2018rvv,Maiani:2016wlq,Zhu:2016arf}.  If $Y(4274)$ is a compact tetraquark state or diquark-antidiquark state~\cite{Chen:2016oma,
Wang:2016dcb,Stancu:2009ka,Agaev:2017foq,Lu:2016cwr,Gui:2018rvv,Maiani:2016wlq,Zhu:2016arf}, we suggest experimental to search for $Y(4274)$ in
the $Y(4274)\to{}\chi_{c0}\eta$ and $Y(4274)\to{}\chi_{c1}\eta$ reactions that the partial decay widths are of the order of 0.0 MeV by assuming $Y(4274)$ 
is a $D_s\bar{D}_{s0}$ molecular state.   Theoretical investigations on decay modes and further experimental information on partial decay widths 
will be helpful to distinguish which the inner structure of the $Y(4274)$ state is possible.

However, a $P-$wave $D_s\bar{D}_{s0}$ molecular with the total decay width is at the order of 1.06 to 1.84 MeV [we marked as $Y^{'}(4274)$] is found.
The predicted decay width seems to suggest that it is possible to observe such a state at Belle or Belle II, e.g., via the inclusive invariant mass
distribution $D\bar{D}^{*}$, which is the largest transition.  On the other hand, its production yields at these experimental setups
remain to be studied.

If the $Y(4274)$ could be a $P-$wave $D_s\bar{D}_{s0}$ molecular.   It naturally leads us to think about whether there are contributions come 
from the light pseudoscalar meson exchanges $(\pi, K, \eta, \eta')$ and the vector meson exchanges $(\rho,\omega, K^*)$..., or there exist other 
decay mechanisms for the few-body systems.   However, only $J/\psi$, $\phi$, $\eta$, $K$, $D$, $D_s$, and $D_s^{*}$ exchange are included because 
there is no information on studies about other meson exchange in the vertices of charm-strange meson.  Moreover, we always think the two-body decay 
modes of the multi-quark states are usually the dominant ones.   Hence, the currently calculation is enough to explain the $Y(4274)$ cannot be a 
$P-$wave $D_s\bar{D}_{s0}$ molecular.

\section*{Acknowledgements}
This work was supported by the National Natural Science Foundation
of China under Grant No.12104076, the Science and Technology
Research Program of Chongqing Municipal Education Commission
(Grant No. KJQN201800510), and the Opened Fund
of the State Key Laboratory on Integrated Optoelectronics
(GrantNo. IOSKL2017KF19).  Yin Huang want to thanks
the support from the Development and Exchange Platform for
the Theoretic Physics of Southwest Jiaotong University under
Grants No.11947404 and No.12047576,  the Fundamental Research
Funds for the Central Universities(Grant No.
2682020CX70), and the National Natural Science Foundation
of China under Grant No.12005177.



\begin{thebibliography}{99}
\bibitem{Zyla:2020zbs}
P.~A.~Zyla \textit{et al.} [Particle Data Group],
PTEP \textbf{2020}, 083C01 (2020).


\bibitem{Aaij:2019vzc}
R.~Aaij \textit{et al.} [LHCb],
Phys. Rev. Lett. \textbf{122}, 222001 (2019).


\bibitem{LHCb:2020jpq}
R.~Aaij \textit{et al.} [LHCb],
Sci. Bull. \textbf{66}, 1278-1287 (2021).

\bibitem{LHCb:2016axx}
R.~Aaij \textit{et al.} [LHCb],
Phys. Rev. Lett. \textbf{118}, 022003 (2017).


\bibitem{LHCb:2016nsl}
R.~Aaij \textit{et al.} [LHCb],
Phys. Rev. D \textbf{95},012002 (2017).



\bibitem{Yi:2010aa}
K.~Yi [CDF],
PoS \textbf{ICHEP2010}, 182 (2010)
arXiv:1010.3470.


\bibitem{Chen:2016oma}
H.~X.~Chen, E.~L.~Cui, W.~Chen, X.~Liu and S.~L.~Zhu,
Eur. Phys. J. C \textbf{77}, 160 (2017).


\bibitem{Wang:2016dcb}
Z.~G.~Wang,
Eur. Phys. J. C \textbf{77}, 174 (2017).


\bibitem{Stancu:2009ka}
F.~Stancu,
J. Phys. G \textbf{37}, 075017 (2010)
[erratum: J. Phys. G \textbf{46}, 019501 (2019)].



\bibitem{Lu:2016cwr}
Q.~F.~L\"u and Y.~B.~Dong,
Phys. Rev. D \textbf{94},074007 (2016).


\bibitem{Gui:2018rvv}
L.~C.~Gui, L.~S.~Lu, Q.~F.~L\"u, X.~H.~Zhong and Q.~Zhao,
Phys. Rev. D \textbf{98}, 016010 (2018).


\bibitem{Agaev:2017foq}
S.~S.~Agaev, K.~Azizi and H.~Sundu,
Phys. Rev. D \textbf{95}, 114003 (2017).


\bibitem{Maiani:2016wlq}
L.~Maiani, A.~D.~Polosa and V.~Riquer,
Phys. Rev. D \textbf{94},  054026 (2016).


\bibitem{Zhu:2016arf}
R.~Zhu,
Phys. Rev. D \textbf{94}, 054009 (2016).


\bibitem{He:2016pfa}
J.~He,
Phys. Rev. D \textbf{95}, 074004 (2017).


\bibitem{Liu:2013rxa}
Y.~R.~Liu,
Phys. Rev. D \textbf{88}, 074008 (2013)
doi:10.1103/PhysRevD.88.074008
[arXiv:1304.7467 [hep-ph]].


\bibitem{Ma:2010xx}
  Y.~L.~Ma,
  Estimates for $X(4350)$ Decays from the Effective Lagrangian Approach,
  Phys.\ Rev.\ D {\bf 82}, 015013 (2010).



\bibitem{Dong:2010gu}
  Y.~Dong, A.~Faessler, T.~Gutsche and V.~E.~Lyubovitskij,
  Phys.\ Rev.\ D {\bf 81}, 074011 (2010).

\bibitem{Faessler:2007gv}
  A.~Faessler, T.~Gutsche, V.~E.~Lyubovitskij and Y.~L.~Ma,
  Phys.\ Rev.\ D {\bf 76}, 014005 (2007).

\bibitem{Faessler:2007us}
  A.~Faessler, T.~Gutsche, V.~E.~Lyubovitskij and Y.~L.~Ma,
  Phys.\ Rev.\ D {\bf 76}, 114008 (2007).

\bibitem{Dong:2008gb}
  Y.~Dong, A.~Faessler, T.~Gutsche and V.~E.~Lyubovitskij,
  Phys.\ Rev.\ D {\bf 77}, 094013 (2008).


\bibitem{Dong:2009uf}
  Y.~Dong, A.~Faessler, T.~Gutsche and V.~E.~Lyubovitskij,
  J.\ Phys.\ G {\bf 38}, 015001 (2011).

\bibitem{Dong:2009yp}
  Y.~Dong, A.~Faessler, T.~Gutsche, S.~Kovalenko and V.~E.~Lyubovitskij,
  Phys.\ Rev.\ D {\bf 79}, 094013 (2009).


\bibitem{Dong:2017rmg}
  Y.~Dong, A.~Faessler, T.~Gutsche, Q.F.~L\"{u} and V.~E.~Lyubovitskij,
  Phys.\ Rev.\ D {\bf 96}, 074027 (2017).


\bibitem{Dong:2014ksa}
  Y.~Dong, A.~Faessler, T.~Gutsche and V.~E.~Lyubovitskij,
  Phys.\ Rev.\ D {\bf 90}, 094001 (2014).

\bibitem{Dong:2014zka}
  Y.~Dong, A.~Faessler, T.~Gutsche and V.~E.~Lyubovitskij,
  Phys.\ Rev.\ D {\bf 90}, 074032 (2014).


\bibitem{Dong:2013kta}
  Y.~Dong, A.~Faessler, T.~Gutsche and V.~E.~Lyubovitskij,
  Phys.\ Rev.\ D {\bf 89}, 034018 (2014).


\bibitem{Dong:2013iqa}
  Y.~Dong, A.~Faessler, T.~Gutsche and V.~E.~Lyubovitskij,
  Phys.\ Rev.\ D {\bf 88}, 014030 (2013).

\bibitem{Dong:2013rsa}
  Y.~Dong, A.~Faessler, T.~Gutsche and V.~E.~Lyubovitskij,
  Few Body Syst.\  {\bf 54}, 1011 (2013).


\bibitem{Dong:2012hc}
  Y.~Dong, A.~Faessler, T.~Gutsche and V.~E.~Lyubovitskij,
  J.\ Phys.\ G {\bf 40}, 015002 (2013).


\bibitem{Dong:2011ys}
  Y.~Dong, A.~Faessler, T.~Gutsche, S.~Kumano and V.~E.~Lyubovitskij,
  Phys.\ Rev.\ D {\bf 83}, 094005 (2011).


\bibitem{Dong:2010xv}
  Y.~Dong, A.~Faessler, T.~Gutsche, S.~Kumano and V.~E.~Lyubovitskij,
  Phys.\ Rev.\ D {\bf 82}, 034035 (2010).


\bibitem{Dong:2009tg}
  Y.~Dong, A.~Faessler, T.~Gutsche and V.~E.~Lyubovitskij,
  Phys.\ Rev.\ D {\bf 81}, 014006 (2010).


\bibitem{Dong:2017gaw}
  Y.~Dong, A.~Faessler and V.~E.~Lyubovitskij,
  Prog.\ Part.\ Nucl.\ Phys.\  {\bf 94}, 282 (2017).


\bibitem{Yang:2021pio}
F.~Yang, Y.~Huang and H.~Q.~Zhu,
[arXiv:2107.13267 [hep-ph]].


\bibitem{Zhu:2020jke}
H.~Zhu, N.~Ma and Y.~Huang,
Eur. Phys. J. C \textbf{80}, 1184 (2020).


\bibitem{Weinberg:1962hj}
  S.~Weinberg,
  Phys.\ Rev.\  {\bf 130}, 776 (1963).

\bibitem{Salam:1962ap}
  A.~Salam,
  Nuovo Cim.\  {\bf 25}, 224 (1962).

\bibitem{Oset:2002sh}
  E.~Oset, J.~R.~Pelaez and L.~Roca,
  Phys.\ Rev.\ D {\bf 67}, 073013 (2003).

\bibitem{Bramon:1992kr}
  A.~Bramon, A.~Grau and G.~Pancheri,
  Phys.\ Lett.\ B {\bf 283}, 416 (1992).


\bibitem{Colangelo:2003sa}
P.~Colangelo, F.~De Fazio and T.~N.~Pham,
Phys. Rev. D \textbf{69}, 054023 (2004).


\bibitem{Casalbuoni:1996pg}
R.~Casalbuoni, A.~Deandrea, N.~Di Bartolomeo, R.~Gatto, F.~Feruglio and G.~Nardulli,
Phys. Rept. \textbf{281}, 145-238 (1997).



\bibitem{Huang:2019qmw}
Y.~Huang, M.~Z.~Liu, Y.~W.~Pan, L.~S.~Geng, A.~Mart\'\i{}nez Torres and K.~P.~Khemchandani,
Phys. Rev. D \textbf{101},014022 (2020).


\bibitem{Gamermann:2006nm}
D.~Gamermann, E.~Oset, D.~Strottman and M.~J.~Vicente Vacas,
Phys. Rev. D \textbf{76}, 074016 (2007).



\bibitem{BaBar:2003oey}
B.~Aubert \textit{et al.} [BaBar],
Phys. Rev. Lett. \textbf{90}, 242001 (2003).



\bibitem{Liu:2006jx}
X.~Liu, Y.~M.~Yu, S.~M.~Zhao and X.~Q.~Li,
Eur. Phys. J. C \textbf{47}, 445-452 (2006).


\bibitem{He:2011ed}
J.~He and X.~Liu,
Eur. Phys. J. C \textbf{72}, 1986 (2012).



\bibitem{BESIII:2017vdm}
M.~Ablikim \textit{et al.} [BESIII],
Phys. Rev. D \textbf{97}, 051103 (2018).


\bibitem{BaBar:2006eep}
B.~Aubert \textit{et al.} [BaBar],
Phys. Rev. D \textbf{74}, 032007 (2006).






\end{thebibliography}
\end{document}